\documentclass[aps,prd,preprint,superscriptaddress,showpacs,floatfix,nobibnotes]{revtex4-1}

\usepackage{latexsym}
\usepackage{amsmath}
\usepackage{amssymb}
\usepackage{graphicx}
\usepackage{enumitem}
\usepackage[usenames]{color}

\usepackage{bm}
\usepackage{epsfig}
\usepackage{subfigure}
\newcommand{\bea}{\begin{eqnarray}}
\newcommand{\eea}{\end{eqnarray}}

\usepackage[section]{placeins}

\begin{document}

\title{The role of frequency dependence in dynamical gap generation in graphene}

\author{M.E. Carrington}
\email[]{carrington@brandonu.ca} \affiliation{Department of Physics, Brandon University, Brandon, Manitoba, R7A 6A9 Canada}\affiliation{Winnipeg Institute for Theoretical Physics, Winnipeg, Manitoba}

\author{C.S. Fischer}
\email[]{} \affiliation{Institut f\"{u}r Theoretische Physik, Justus-Liebig-Universit\"{a}t Giessen, Heinrich-Buff-Ring 16, 35392 Giessen, Germany}

\author{L. von Smekal}
\email[]{} \affiliation{Institut f\"{u}r Theoretische Physik, Justus-Liebig-Universit\"{a}t Giessen, Heinrich-Buff-Ring 16, 35392 Giessen, Germany}

\author{M.H. Thoma}
\email[]{} \affiliation{I. Physikalisches Institut, Justus-Liebig-Universit\"{a}t Giessen, Heinrich-Buff-Ring 16, 35392 Giessen, Germany}

\date{\today}

\begin{abstract}
  We study the frequency dependencies of the fermion and photon dressing functions in dynamical gap generation in graphene. 
  We use a low energy effective QED-like description, but within this approximation, we include all frequency dependent 
  effects including retardation. We obtain the critical coupling by calculating the gap using a non-perturbative Dyson-Schwinger 
  approach. Compared to the results of our previous calculation [M.E. Carrington et al., Phys. Rev. B {\bf94}, 125102 (2016)] which used a Lindhard screening approximation 
  instead of including a self-consistently calculated dynamical screening function, the critical coupling is substantially reduced. 
\end{abstract}


\normalsize
\maketitle

\normalsize

\section{Introduction}
\label{introduction-section}

Graphene is a 2-dimensional crystal of carbon atoms with many possible applications in a wide range of technological fields. 
Theoretically graphene is interesting to physicists, in part, because it provides a  condensed matter analogue of many problems that are studied in particle physics using relativistic quantum field theory. 
Two recent reviews of the properties of graphene are Refs.~\cite{neto1,neto2}.

Graphene is normally found in a (semi) metal state, but if quasi-particle interactions are strong enough they could produce a gap and cause a phase transition to an insulating state. 
The effective coupling can be written as 
$\alpha = \frac{e^2}{4\pi \epsilon \hbar v_F}$
where 
$v_F\sim c/300$ is the velocity of a massless electron in graphene and  
$\epsilon\ge 1$ is related to the screening properties of the graphene sheet.
The maximum possible effective coupling is the vacuum value (corresponding to $\epsilon=1$) and is about $\alpha_{\rm max} = 2.2$. 
To determine theoretically if a gap is formed in the physical system, we calculate the critical coupling $\alpha_c$ for gap formation. If $\alpha_c > 2.2$, then the physical interactions are not strong enough to produce an insulating phase. 
Experiments indicate that the insulating state is not physically realizable \cite{elias11}.

Many theoretical calculations of the critical coupling for gap formation have been done. 
Some earlier calculations can be found in Refs. \cite{khv,brane,liu,gor,case,gonz,drut,hands,pop,sols,rong-liu,teber}. 
There are two major issues that are difficult to correctly formulate in a theoretical calculation: screening and frequency dependent effects. 
Realistically screened Coulomb interactions have been included by using a constrained random phase approximation in Ref.~\cite{wehling11}, and in hybrid-Monte-Carlo simulations on a hexagonal lattice \cite{ulybyshev13,lorenz1}. However, these calculations are done within the Coulomb approximation and therefore ignore many frequency dependent effects. 

In this paper we are interested in the second issue - the influences of frequency dependent screening and retardation effects.
We consider the simplest form, mono-layer graphene, in which the carbon atoms are arranged in a 2-dimensional hexagonal lattice, and we work at half filling (zero chemical potential).
The low energy dynamics are described by a continuum quantum field theory
in which the electronic quasi-particles have a linear Dirac-like dispersion relation of the form $E=\pm v_F p$. 
We therefore use an effective QED-like description for the low-energy excitations which allows us to correctly include all frequency effects, but does not allow for the inclusion of screening from the $\sigma$-band electrons and localized higher energy states. 
We include non-perturbative effects by introducing fermion and photon dressing functions, and solving a set of coupled Dyson-Schwinger (DS) equations. Using this formulation, a gap function, which would produce a phase transition to an insulating state, would appear as a dynamically generated fermion mass. 

The DS equations are an infinite hierarchy of integral equations that must be decoupled by introducting some additional approximation. In our work, we perform this decoupling by using an approximation for the vertex which is given in equation (\ref{vert-def}). 
Our approximation corresponds to the first term in the Ball-Chiu vertex \cite{ball-chu}. 
The Ball-Chiu vertex is the most general form of the vertex that satisfies gauge invariance, and the first term of it is the easiest piece to calculate. 
In a previous paper \cite{giessen1}, we solved the fermion Dyson-Schwinger equation using the full Ball-Chiu vertex and a perturbative Lindhard-type screening function in the photon propagator, including retardation effects, and using a variety of ans\" atze for the vertex function. 
Our results indicate that the role of frequency dependencies in the photon propagator and fermion dressing functions is important, but the calculation is relatively insensitive to the form of the vertex function. 

\begin{figure}[!htp]
\centering
\mbox{\subfigure[]{\includegraphics[width=6.2in]{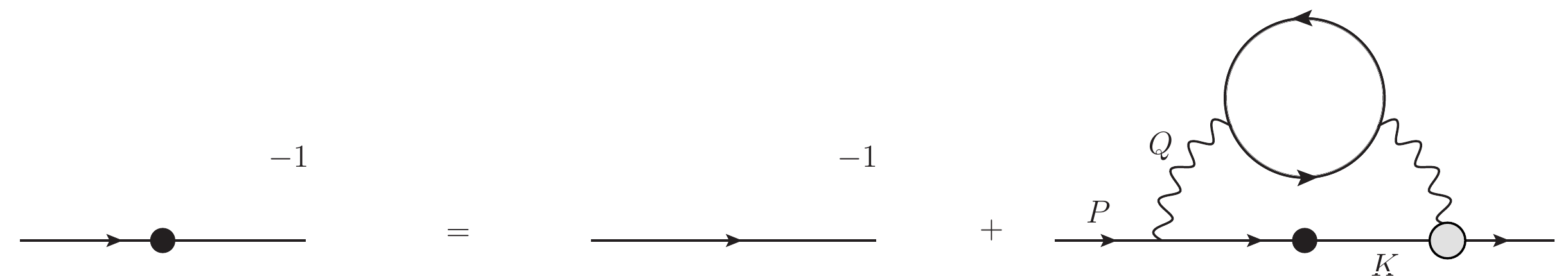}}}\\
\mbox{\subfigure[]{\includegraphics[width=6.2in]{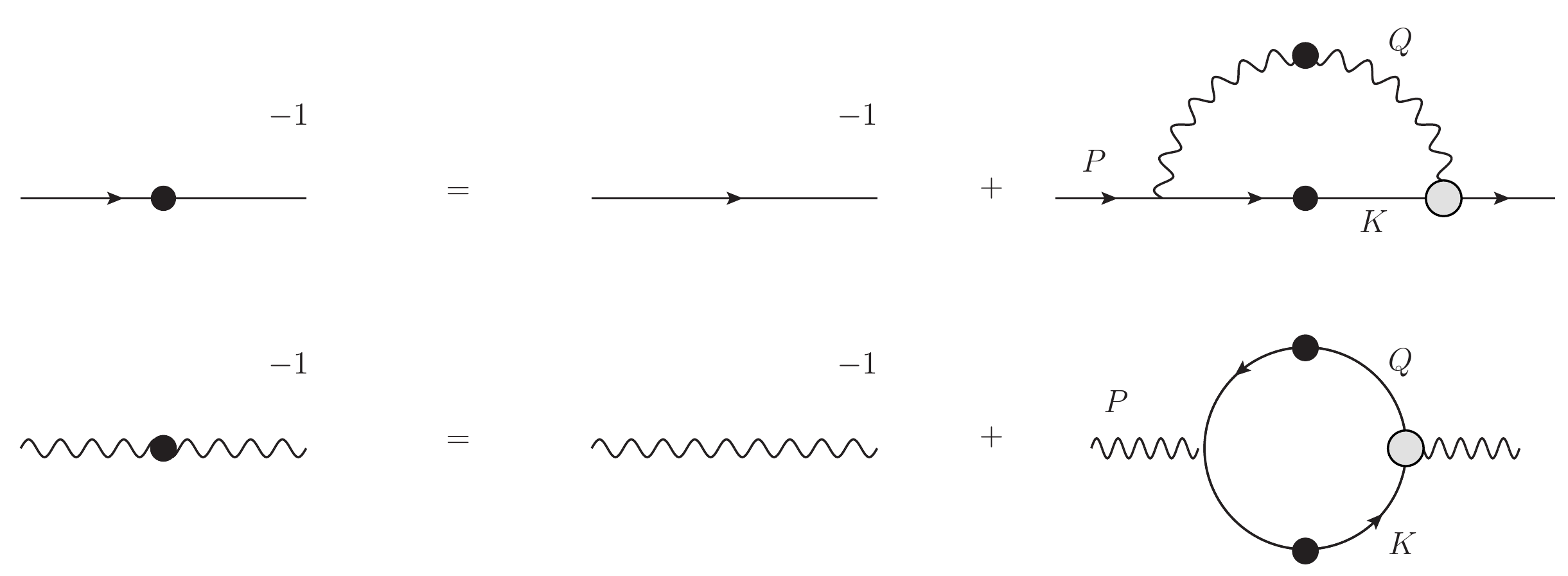}}}
\caption{Graphical representation of some DS equations. Solid dots represent dressed propagators and the gray dots are  model dependent vertices. In part (a) we show the equation solved in our previous paper \cite{giessen1}, in which a 1 loop approximation for the photon self energy and the Ball-Chiu vertex are used. In part (b) we show the set of self-consistent equations that we solve in this paper, where the vertex represented by the gray dot is given in equation (\ref{vert-def}). \label{SD-fig}}
\end{figure}

The potentially important simplifying assumption in our previous calculation was the use of the Lindhard screening function in the photon propagator. This screening function is a specific approximation to the electric part of the one loop vacuum photon polarization tensor. The approximation is usually justified by the idea that the vanishing fermion density of states at the Dirac points indicates that the
one loop contribution to the photon polarization should dominate (see, for example, \cite{neto2,stauber,dietz}). 
However, our previous calculations \cite{pop, giessen1} show clearly that it is crucial to include without approximation the fermion dressing functions that give the renormalisation of the fermi velocity. Since the photon polarization is determined from a fermion loop, we therefore expect that a self-consistent calculation of photon screening could have a significant effect on the result.
In this paper, we present the complete version of our previous calculation which does not make use of the Lindhard screening approximation, but instead includes a self-consistently determined photon dressing function. 
For comparison, we also consider a self-consistent Coulomb approximation, which includes a self-consistent photon dressing function but neglects retardation effects. 
A diagrammatic representation of the approximations discussed above is shown in Fig. \ref{SD-fig}. 

We comment that the Ball-Chiu vertex ansatz that we use represents a first step towards a full calculation that would involve self-consistent 3 point functions. 
The ansatz is modelled on gauge invariance, and experience from many previous calculations with 2+1 dimensional QED. 
It allows us to obtain and solve a complete and closed set of integral equations for the 2 point functions of the theory. 
A disadvantage of such a vertex ansatz is that the resulting integral equations cannot be directly related to Feynman diagrams, and there is no straightforward way to estimate corrections. 
However, calculations involving self-consistent vertices are computationally much more difficult, and the fact that results seem insensitive to the form of the vertex ansatz may indicate that they are unnecessary.

Our results show that there is in fact a significant change in the critical coupling when the photon self energy is calculated self-consistently. Relative to the Lindhard calculation in \cite{giessen1}, the critical coupling is reduced by about 10\%. When retardation effects are neglected, the self-consistent result is reduced by approximately 9\%. 
We emphasize that the precise numerical values of the critical couplings that we obtain are not meant to be realistic, 
since they will clearly be changed (in a predictable manner) by short distance screening effects which we have not included 
(additional screening would increase the critical coupling that we obtain). 
However, our calculation provides valuable information about the extent to which the Lindhard and Coulomb approximations are valid.

\section{Notation}
\label{notation-section}

The Euclidean action of the low energy effective theory is given by
\bea
\label{action}
S=\int d^3 x \sum_{a}\bar\psi_a \left(i\partial_\mu -e A_\mu\right)M_{\mu\nu}\gamma_\nu \psi _a - \frac{\epsilon}{4e^2}\int d^3x F_{\mu\nu}\frac{1}{2\sqrt{-\partial^2}}F_{\mu\nu} \text{ + gauge fixing}
\eea
where the Greek indices take values $\in\{0,1,2\}$. 
The gauge field action is non-local because the photon which mediates the interactions between the electrons propagates in the 3+1 dimensional space-time, and therefore out of the graphene plane.
This ``brane action'' can be derived by integrating out the photon momentum modes in the third spatial dimension, and has some significant differences from 3-dimensional QED \cite{brane,marino}. 
The fermionic part of the action looks like a free Dirac theory with a linear dispersion relation, and this is a reflection of the fact that the low energy effective theory is a good description of the system close to the Dirac points. 
Quasi-particle excitations on both sub-lattices, with momenta close to either of the two Dirac points, are represented by four-component Dirac spinors.
The true spin of the electrons appears formally  as an additional flavour
quantum number, and we take $N_f = 2$ for monolayer graphene. The three 4-dimensional
$\gamma$-matrices form a reducible representation of the Clifford algebra $\{\gamma_\mu,\gamma_\nu\} = 2\delta_{\mu\nu}$ in 2+1 dimensions.
The matrix  $M$ is defined
\bea
\label{Mdef}
M = 
\left[\begin{array}{ccc}
~1~ & ~0~ & ~0~ \\
0 & v_F  & 0 \\
0 & 0 & v_F   \\
\end{array}
\right]\,,
\eea
and we note that Lorentz invariance is explicitly broken by taking $v_F\ne 1$. 

The Euclidean space Feynman rules, in Landau gauge, are 
\bea
\label{bareFR}
&& S^{(0)}(P) = -\big[i\gamma_\mu M_{\mu\nu} P_\nu\big]^{-1}\,,\\[2mm]
&& G^{(0)}_{\mu\nu}(P)=\big[\delta_{\mu\nu}-\frac{P_\mu P_\nu}{P^2}\big]\,\frac{1}{2\sqrt{P^2}}\,, \\[1mm]
\label{barevert}
&& \Gamma^{(0)}_\mu = M_{\mu\nu}\gamma_\nu\,,
\eea
where we write $P_\mu = (p_0,\vec p)$ and $P^2=p_0^2+p^2$, and similarly for the momenta $K$ and $Q=K-P$.
We also use the notation
\bea
\label{d4K}
dK = \int \frac{dk_0\,d^2k}{(2\pi)^3}\,.
\eea

The full fermion propagator has three dressing functions because of the broken Lorentz invariance. 
We call these three functions $Z(p_0,\vec p)$, $A(p_0,\vec p)$ and $\Delta(p_0,\vec p)$. Defining the diagonal 3$\times$3 matrix
\bea
\label{Amatrix}
H(p_0,\vec p) = 
\left[\begin{array}{ccc}
Z(p_0,\vec p) & 0 & 0 \\
0 & v_F A(p_0,\vec p)  & 0 \\
0 & 0 & v_F A(p_0,\vec p)   
\end{array}
\right]\,,
\eea
the dressed fermion propagator is written
\bea
\label{SF}
&& S^{-1}(P) = - i \gamma_\mu H_{\mu\tau}(p_0,\vec p) P_\tau + \Delta(p_0,\vec p)\,,
\eea
and the inverse propagator takes the form 
\bea
\label{SF2}
&& S^{-1}(P) = (S^{(0)})^{-1}(P)+\Sigma(P)\,.
\eea 
The DS equation for the fermion self energy is
\bea
\label{fermion-SD}
&& \Sigma(p_0,\vec p) = e^2\int dK \,G_{\mu\nu}(q_0,\vec q)\,M_{\mu\tau}\,\gamma_\tau \, S(k_0,\vec k) \,\Gamma_\nu\,.
\eea
We define projection operators and decompose the polarization tensor,
\bea
&& P^1_{\mu\nu}=\delta_{\mu\nu}-\frac{P_\mu P_\nu}{P^2}\,,~~ P^3_{\mu\nu} = \frac{n_\mu n_\nu}{n^2}\,,\\
&& P^2_{\mu\nu} = \frac{P_\mu P_\nu}{P^2}\,,~~ \hat P^4_{\mu\nu} =(n_\mu P_\nu+n_\nu P_\mu)/p\,, \\ 
&& n_\mu = \delta_{\mu 0}-\frac{p_0 P_\mu}{P^2}\,,\\
&& \Pi_{\mu\nu}=\alpha(p_0,p) P^1_{\mu\nu}+ \beta(p_0,p) P^2_{\mu\nu}+ \gamma(p_0,p) P^3_{\mu\nu} + \hat\delta(p_0,p) \hat P^4_{\mu\nu}\,.
\eea 
The inverse photon propagator in Lorentz gauge is 
\bea
G_{\mu\nu}^{-1} = \frac{2}{\sqrt{P^2}}P^2\big(P_{\mu\nu}^1+\frac{1}{\xi} P^2_{\mu\nu}\big)+\Pi_{\mu\nu}\,.
\eea
Choosing Landau gauge ($\xi=0$) and taking the polarization tensor to be transverse, the propagator is 
\bea
\label{fullG}
&& G_{\mu\nu} =  
\frac{P_{\mu\nu}^1}{G_T(p_0,\vec p)}+ P_{\mu\nu}^3\left(\frac{1}{G_L(p_0,\vec p)}-\frac{1}{G_T(p_0,\vec p)}\right)\,,\\[4mm]
\label{fullPi}
&& G_T(p_0,\vec p) = 2\sqrt{P^2}+\alpha(p_0,p)\,,~~~G_L(p_0,\vec p) = 2\sqrt{P^2}+\alpha(p_0,p)+\gamma(p_0,p) \,.
\eea
The two propagator components $G_T(p_0,\vec p)$ and $G_L(p_0,\vec p)$ are transverse and longitudinal with respect to the three momentum $\vec p$, and the DS equation for the polarization tensor is
\bea
\label{photon-SD}
&& \Pi_{\mu\nu}(p_0,\vec p) = -e^2\int dK \,{\rm Tr}\,\big[S(q_0,\vec q) \, M_{\mu\tau} \, \gamma_\tau \, S(k_0,\vec k)\, \Gamma_\nu\big]\,.
\eea

To truncate the hierarchy of DS equations, we must choose an ansatz for the vertex $\Gamma$ in equations (\ref{fermion-SD}, \ref{photon-SD}). In \cite{giessen1} we used a non-covariant extension of the Ball-Chiu vertex \cite{ball-chu} which satisfies the Ward identity 
$-i Q_\mu\Gamma_\mu(P,K) = S^{-1}(k_0,\vec k) - S^{-1}(p_0,\vec p)$ and is multiplicatively renormalisable in Landau gauge. 
This vertex is difficult to work with numerically because the integrands contain terms that approach 0/0 $\to$ constant as $K\to P$.
In \cite{giessen1} we found that using the first term in the non-covariant Ball-Chiu vertex (1BC), which is numerically much easier to work with,  produces a result for the critical coupling that agrees with the result from the full vertex to within 0.2\%. In this paper we therefore use the truncated expression
\bea
\label{vert-def}
\Gamma_\mu && = \frac{1}{2}\big(H_{\mu\nu}(p_0,\vec p)+H_{\mu\nu}(k_0,\vec k)\big)\gamma_\nu \,.
\eea
Within this approximation, the only component of the propagator (\ref{fullG}) that contributes is the piece $G_L$ and therefore we only need to calculate one component of the polarization tensor which we write as $\Pi_{00}(p_0,p) = \frac{p^2}{P^2}\big(\alpha(p_0,p) + \gamma(p_0,p)\big)$. 

Using (\ref{vert-def}) it is straightforward to obtain integral expressions for the dressing functions $Z(p_0,p)$, $A(p_0,p)$, $D(p_0,p)$,  and $\Pi_{00}(p_0,p)$ by taking the appropriate projections of the corresponding Dyson-Schwinger equation. The resulting equations are
\bea
\label{Zeqn}
&& Z_p = 1-\frac{2\alpha\pi v_F}{p_0 } \int dK \frac{k_0 q^2 Z_k (Z_p+Z_k)}{Q^2\,G_L S_k} \,,\\
\label{Aeqn}
&& A_p = 1 + \frac{2\alpha\pi v_F}{p^2} \int dK \; \frac{q^2 A_k (Z_p+Z_k) \vec k \cdot \vec p  +  k_0 q_0 Z_k(Z_p+Z_k + A_p+A_k) \vec p \cdot \vec q }{Q^2\,G_L S_k}\,,\\
\label{Deqn}
&&\Delta_p =  2\alpha\pi v_F \int dK \frac{q^2 \Delta _k (Z_p+Z_k)}{Q^2\,G_L S_k}\,, \\[4mm]
\label{Leqn}
&& \Pi_{00}(p_0,p) = -16 \pi v_F \alpha \int \frac{dK}{S_k S_q} \,\left(Z_k+Z_q\right) \left(A_k A_q v_F^2 (\vec k \cdot \vec q)+\Delta _k \Delta _q-k_0 q_0 Z_k
   Z_q\right)\,.
\eea
We note that some of the factors of $Z$ and $A$ on the right sides of equations (20-23) come from the vertex functions. Using the ansatz in equation (19), the dressed vertex has factors of fermion self energy dressing functions from the incoming and outgoing fermion legs of the vertex. For the loop diagram in the first line of figure 1(b) these legs have momenta $K$ and $P$, and therefore the vertex can contribute factors $Z_p$, $A_p$, $Z_k$ and $A_k$ to the right side of equations (20-22).
For the loop diagram in the second line of figure 1(b) the fermion legs of the dressed vertex have momenta $K$ and $Q$, and therefore the vertex can contribute factors $Z_k$, $A_k$, $Z_q$ and $A_q$ on the right side of equation (23).

For comparison, we also perform the calculation with retardation effects neglected, which we call the Coulomb approximation. 
The physical basis of the approximation is the fact that photons move faster than electrons by a factor $1/v_F \sim 300$, which implies that we can take $p_0/p \ll 1$ in the photon propagator. This means that in equations (\ref{fullG}, \ref{fullPi}) we take
\bea
&& \frac{P^3_{\mu\nu}}{G_L} \to \delta_{\mu 0}\delta_{\nu 0}G_{00} \,,~~~~ G_{00} = \frac{1}{f(2\sqrt{f p^2}+ f\Pi_{00})}\bigg|_{f=1}\,,
\eea
resulting in three simplifications of equations (\ref{Zeqn} - \ref{Deqn}):
\begin{enumerate}
\item the second term in the integrand of (\ref{Aeqn}) is no longer present;
\item in each integrand the overall factor $q^2/Q^2 \rightarrow 1$;
\item in the denominator the factor $G_L=2\sqrt{Q^2} + \Pi_L \rightarrow 2q+\Pi_{00}(q_0,\vec q)$.
\end{enumerate}

In all calculations we introduce the same ultraviolet cut-off $\Lambda$ 
in the $k_0$ and $k$ momentum integrals and use a logarithmic scale, to increase the sensitivity of the numerical integration procedure to the infrared regime where the dressing 
functions change most rapidly.
We define dimensionless variables $\hat k_0=k_0/\Lambda$, $\hat p_0=p_0/\Lambda$, $\hat k=k/\Lambda$, $\hat p=p/\Lambda$ 
and $\hat\Delta =\Delta/\Lambda$. The hatted frequency and momentum variables range from zero to one. 
From this point on, we suppress all hats.  

\section{Results}
\label{results-section}

\begin{figure}[!t]
\begin{center}
\includegraphics[width=16cm]{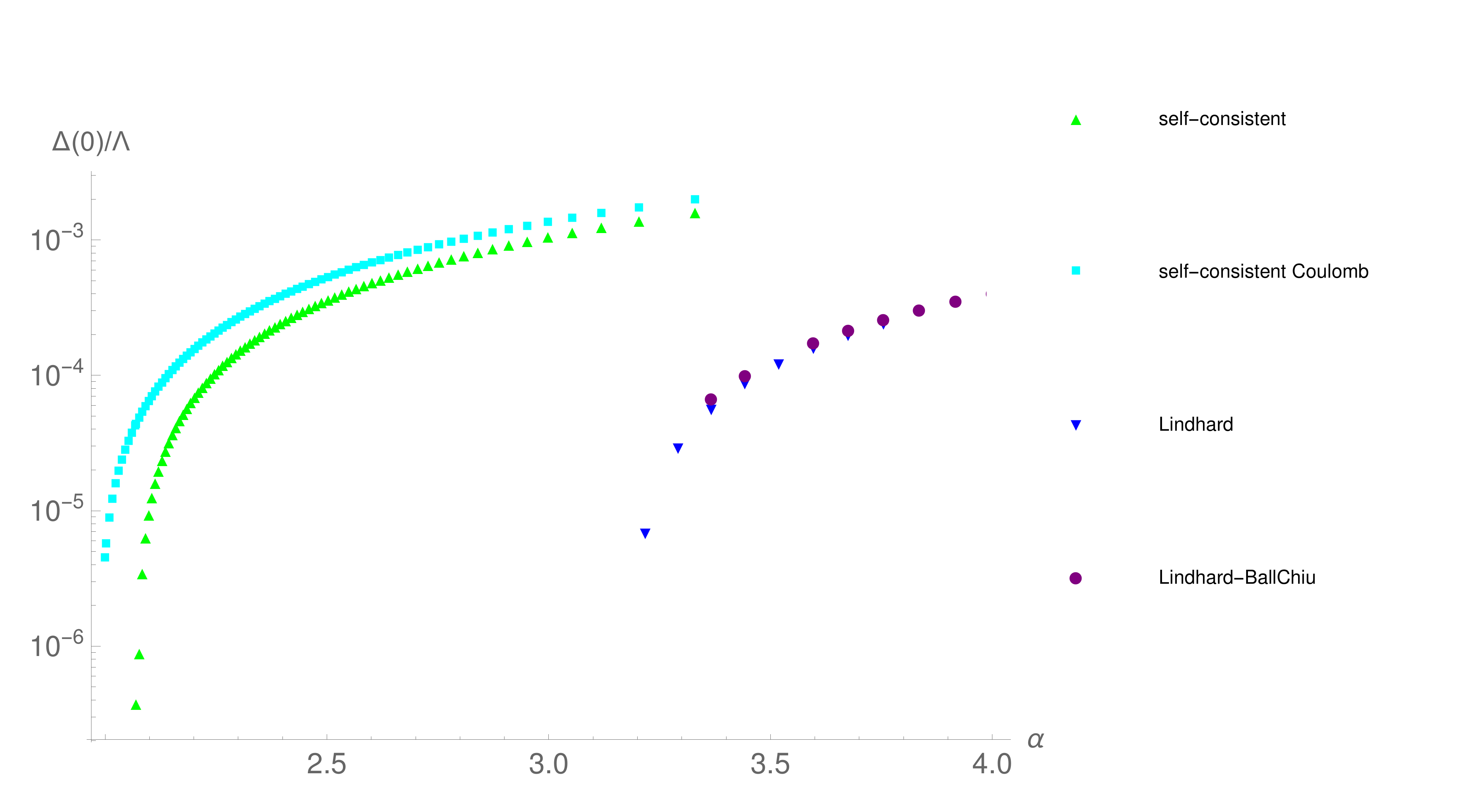}
\end{center}
\caption{The condensate as a function of the coupling\label{alphaC-fig}.}
\end{figure}

\begin{table}[b]
\begin{center}
\begin{tabular}{ |c|c| } 
 \hline
~~~~~~~~~~~ calculation ~~~~~~~~~~~ & ~~~~~~~ $\alpha_c$ ~~~~~~~  \\ 
 \hline
self-consistent & 2.06 \\
self-consistent Coulomb & 1.99 \\
Lindhard & 3.19 \\
  \hline
\end{tabular}
\end{center}
\caption{Results for critical values of the coupling $\alpha$. The first two lines are the results of the present paper, and the result in the last line was obtained in our previous paper \cite{giessen1}. \label{table-alphac}}
\end{table}
In Fig. \ref{alphaC-fig} we show the value of the gap $\Delta(0,0)$ versus $\alpha$. For comparison we have also shown the result obtained in our previous paper using the Lindhard screening function, with both the full Ball Chiu vertex function, and the 1BC-approximation used mainly in this paper (see equation (\ref{vert-def})). As explained in the text above equation (\ref{vert-def}), the truncation of the vertex that we use has almost no effect on the value of the critical coupling. 
We fit the data shown in Fig. \ref{alphaC-fig} using {\it Mathematica}, using three different methods: Spline, Hermite, and Automatic. 
The resulting function is extrapolated to obtain the value of the critical coupling for which the gap goes to zero.  The numbers obtained 
from the three different methods are the same to three decimal places. Our results are collected in Table \ref{table-alphac} 
(the Lindhard result is taken from \cite{giessen1}). The main result of this work is a substantial reduction of the critical 
coupling once the Lindhard approximation is given up in favour of a full self-consistent calculation. Self-consistency in frequencies
is also much more important than the relativistic setup: the self-consistent Coulomb-approximation deviates only mildly from the 
full self-consistent result. This small reduction is consistent with what was found in our previous work \cite{giessen1}.

\begin{figure}[!htp]
\centering
\mbox{\subfigure[]{\includegraphics[width=3.2in]{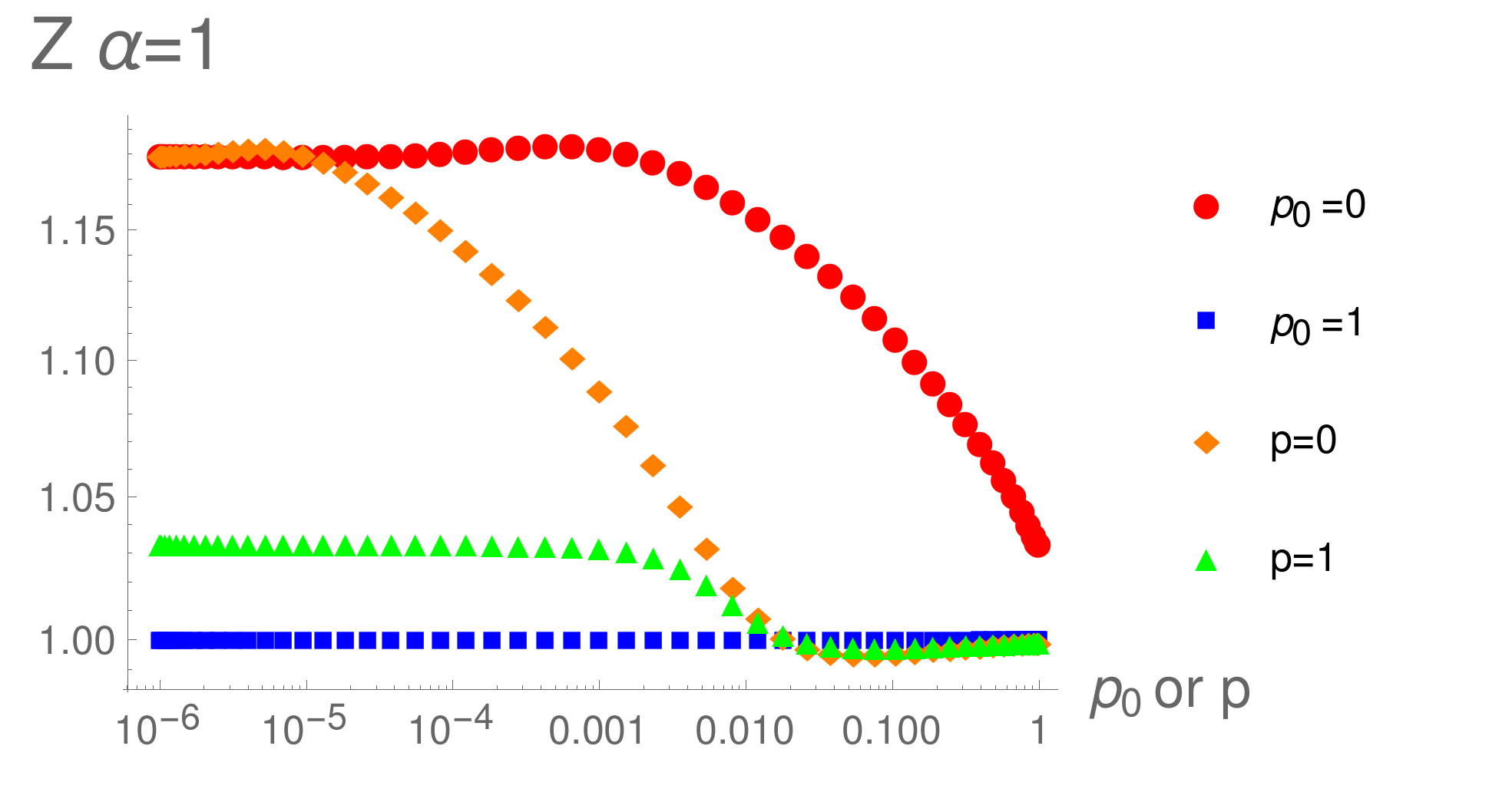}}\quad
\subfigure[]{\includegraphics[width=3.2in]{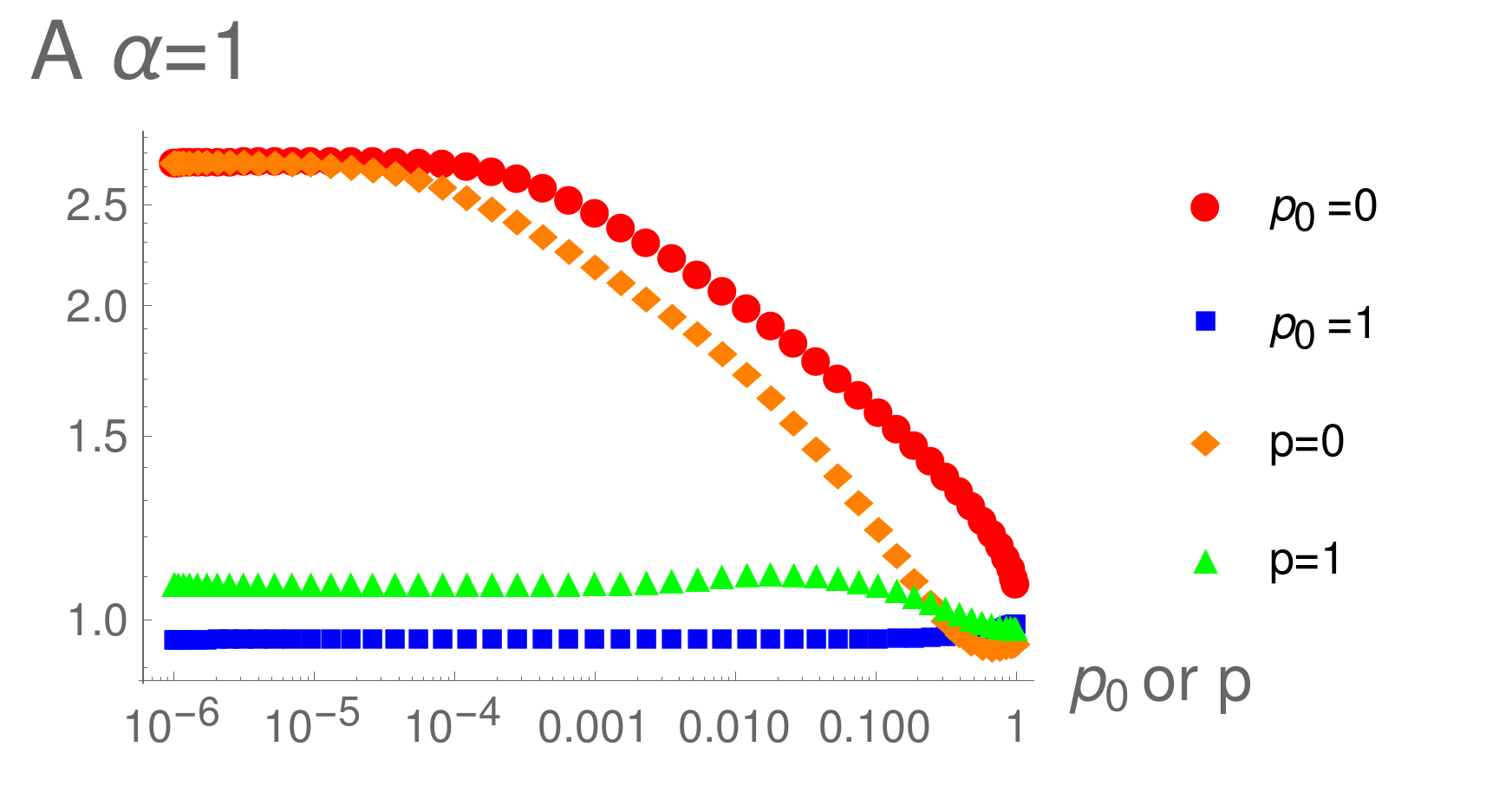} }}
\mbox{\subfigure[]{\includegraphics[width=3.2in]{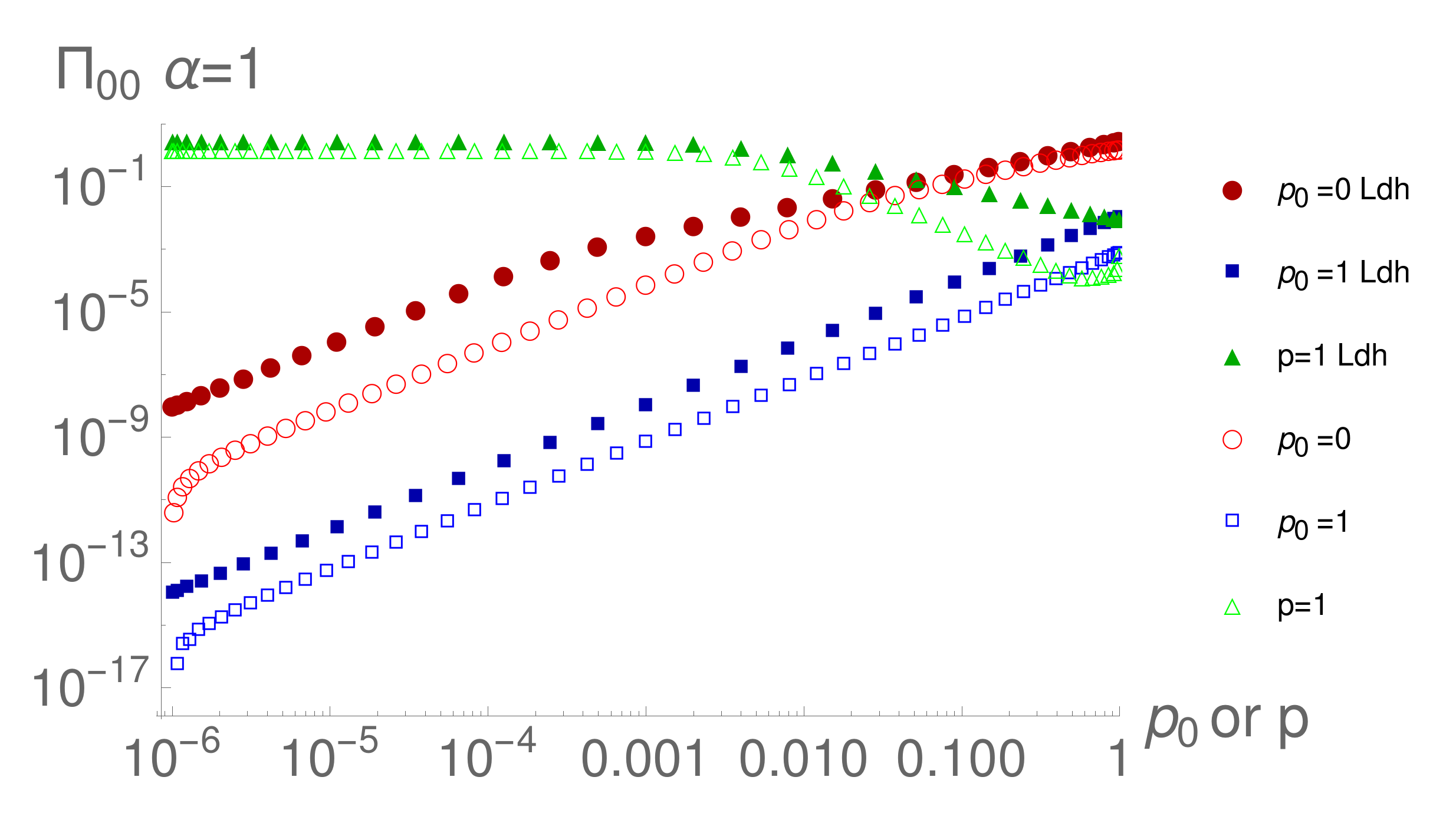}}}
\caption{Momentum dependence of the $Z$, $A$ and $\Pi_{00}$ dressing functions with $\alpha$=1. \label{ZAL-fig}}
\centering
\mbox{\subfigure[]{\includegraphics[width=3.2in]{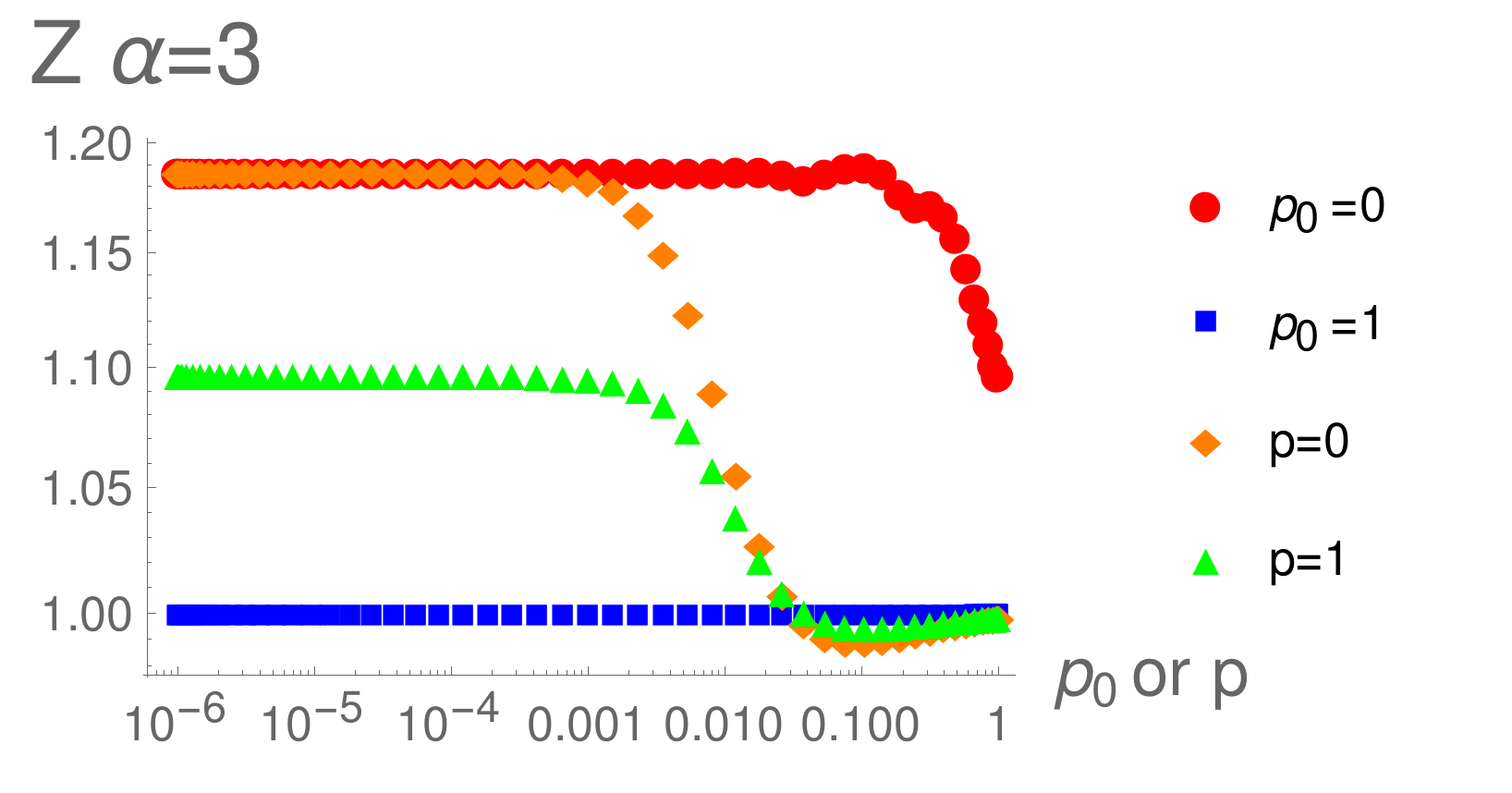}}\quad
\subfigure[]{\includegraphics[width=3.2in]{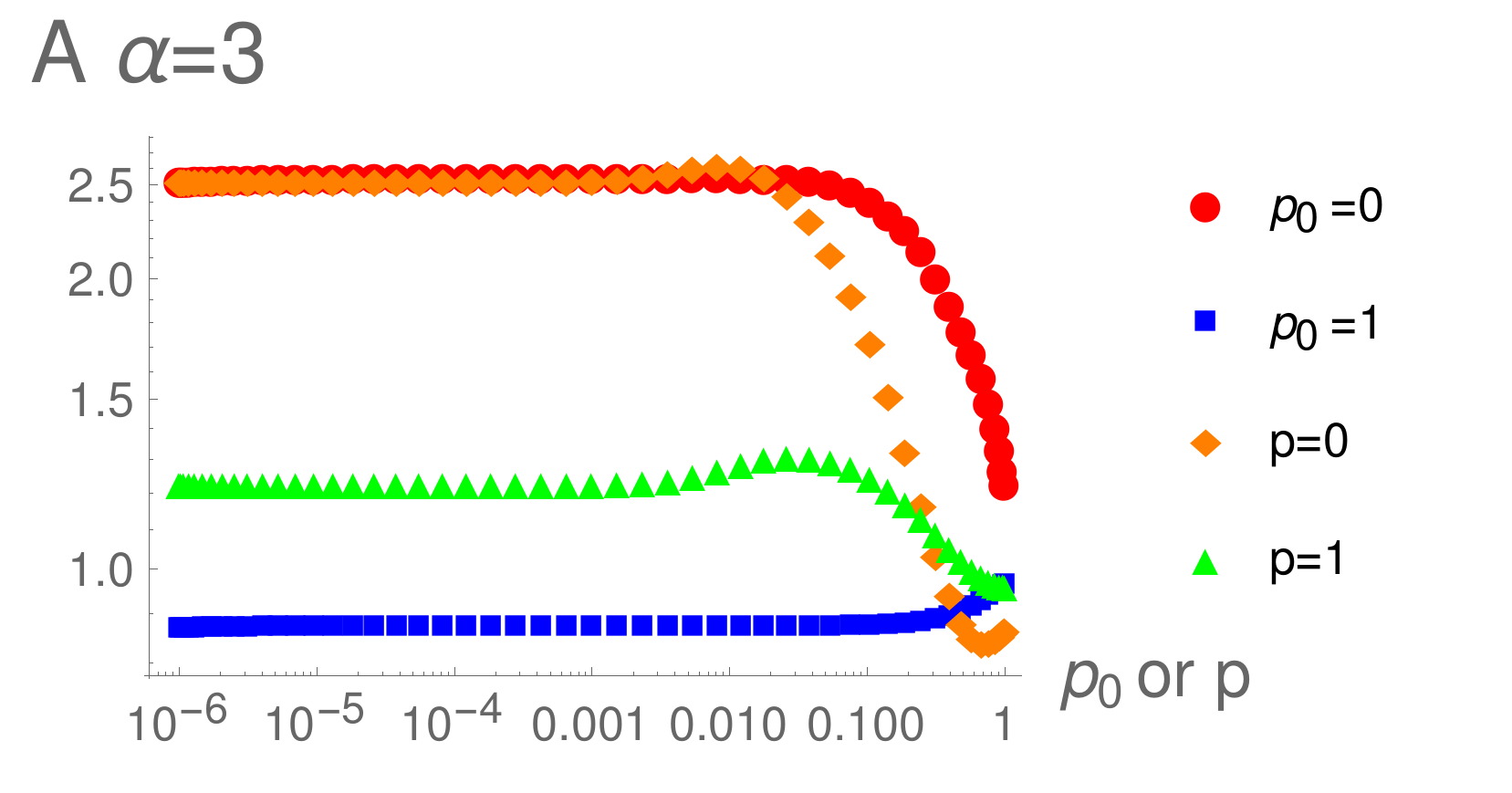} }}
\mbox{\subfigure[]{\includegraphics[width=3.2in]{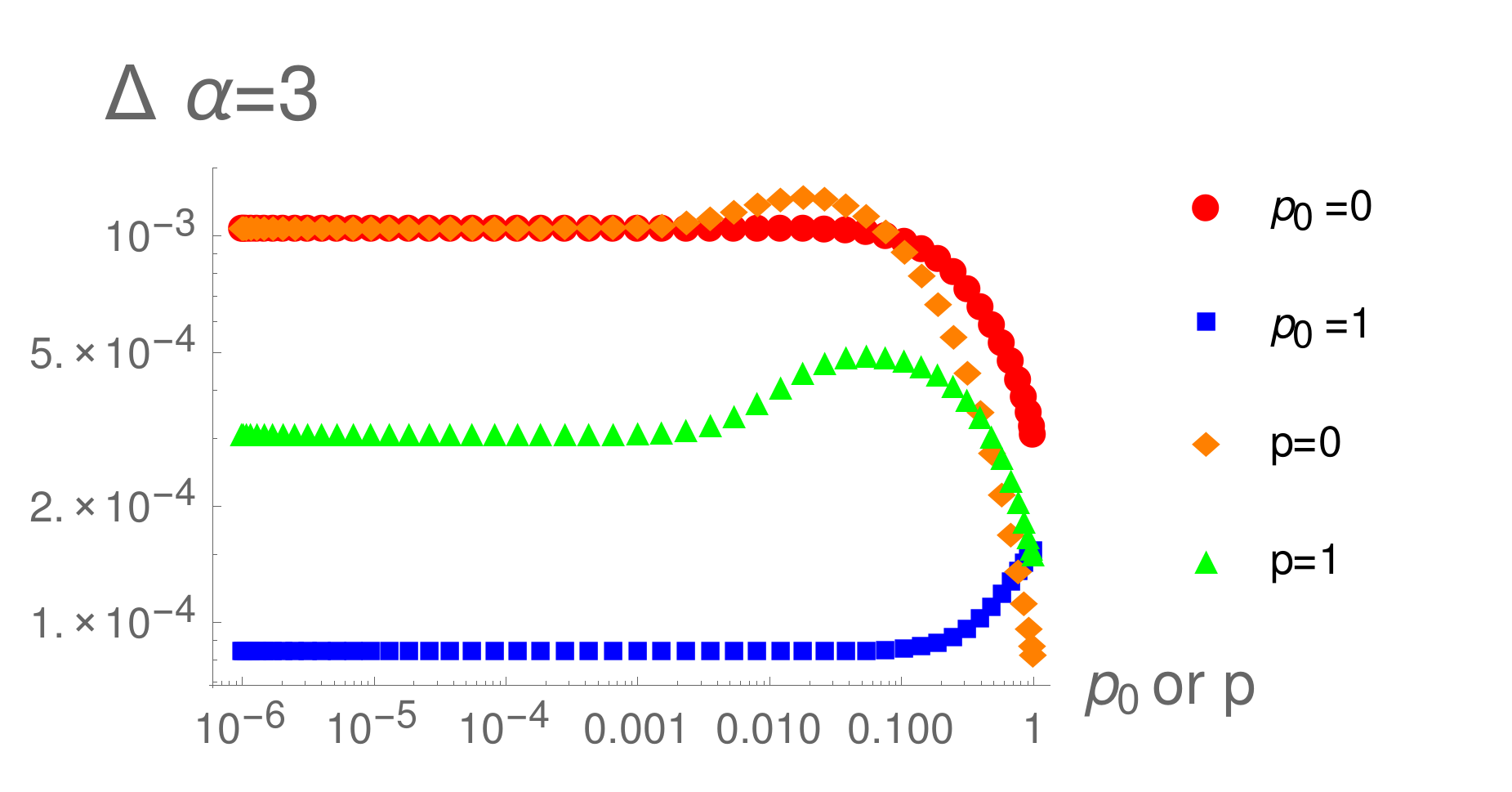}}}
\caption{Momentum dependence of the $Z$, $A$ and $\Delta$  dressing functions with $\alpha=3$. \label{ZADL-fig}}
\end{figure}

To understand the drastic changes induced by self-consistency we discuss our results for the dressing functions for the fermions
and the photon. 
%
In Fig. \ref{ZAL-fig} we show the  momentum dependence of the dressing functions for $(\alpha=1)<\alpha_c$ in the symmetric
phase and in Fig. \ref{ZADL-fig} we show the dressing functions for $(\alpha=3)>\alpha_c$ in the gaped phase.  
Each dressing function is plotted either versus $p$ or $p_0$. 
The variable which is not plotted is held fixed and chosen to be either the smallest or largest value available, which are $10^{-6}$ and 1. 
For $\alpha=1$ there are no plots of $\Delta$ since it is zero below the critical coupling. 
The photon self energy is renormalised numerically by subtracting $\Pi_{00}(p_0,0)$ for each value of $p_0$. 
In Fig. \ref{ZAL-fig}c we show the self-consistent photon dressing function and the Lindhard screening function $ \Pi_{00} = \frac{\pi  \alpha  p^2 v_F}{\sqrt{p^2 v_F^2+p_0^2}}$. We see that the Lindhard screening function is larger than the self-consistently calculated photon self energy.
This result is expected since the photon self energy is calculated from a 1 loop diagram with two fermion propagators, and the Lindhard function uses two bare propagators while the self-consistent fermion dressing functions are consistently greater than one over the full momentum range (see Figs. \ref{ZAL-fig}, \ref{ZADL-fig}). 
 The Lindhard approximation therefore includes an artificially large dynamical screening effect, and consequently produces a larger critical coupling (as seen in Fig. \ref{alphaC-fig} and Table \ref{table-alphac}). 

\begin{figure}[!htp]
\begin{center}
\includegraphics[width=10cm]{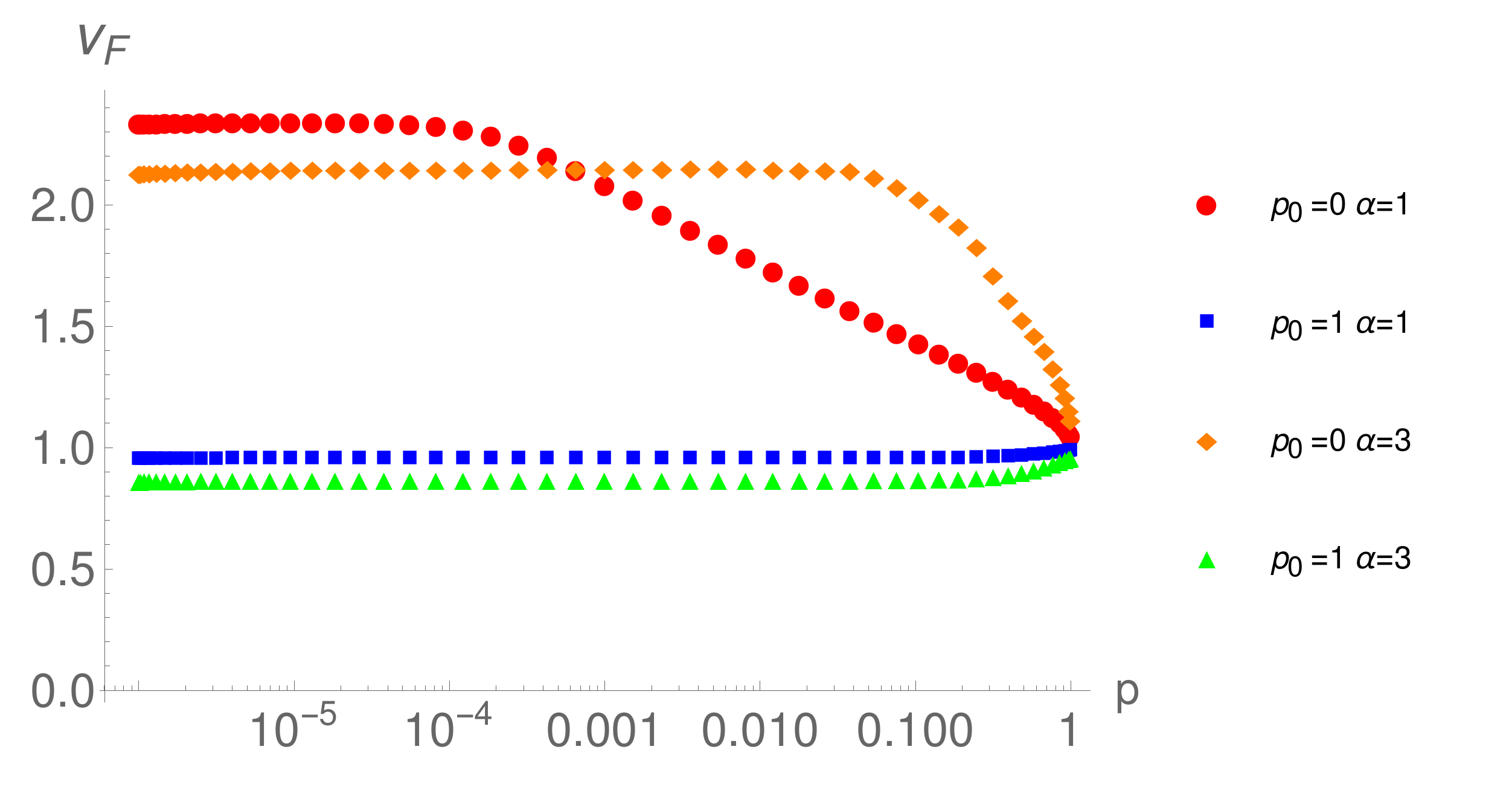}
\end{center}
\caption{ Renormalised fermion velocity\label{vF-fig}}
\end{figure}

In Fig. \ref{vF-fig} we show the renormalised fermion velocity, which is defined as $A(p_0,p)/Z(p_0,p)$. 
The experimentally observed increase in the Fermi velocity below the critical coupling at small frequencies is clearly seen, but suppressed relative to the results of \cite{giessen1}.

\section{Conclusions}

We have reported results of a calculation of the dynamically generated gap using the Dyson-Schwinger equations of 
a low energy effective field theory, which describes some features of mono-layer suspended graphene.
%
In the calculation we have taken into account previously neglected frequency dependencies in a self-consistent way.
Our results show that the inclusion of a self-consistently determined photon self energy substantially reduces the critical coupling, 
relative to that which is obtained with the previously used Lindhard screening function. This result agrees with naive expectations, based on the 
large size of the fermion dressing functions at small momentum. Neglecting retardation effects in the self-consistent calculation 
reduces the critical coupling further, but only by a smaller amount. 

We remind the reader that the main goal of our work was to study the effect of 
frequency dependencies on the critical coupling using an effective low energy theory. 
The precise numerical values of the critical couplings that we obtain are not meant to be realistic, since they will clearly 
be changed (in a predictable manner) by short distance screening effects which we have not included.
Our results provide valuable information about the validity of the frequency approximations that are commonly used in 
calculations done on honeycomb lattices.

\section{Acknowledgements}
This work has been supported by the Natural Sciences and Engineering Research Council of Canada, 
the Helmholtz International Center for FAIR within the LOEWE program of the State of Hesse, 
and by the DFG research grant SM 70/3-1.

\end{document}